\documentclass[prb,aps,twocolumn,showpacs,10pt]{revtex4-1}
\usepackage{amsmath,amssymb,bm,graphicx,dsfont,color}

\usepackage[latin9]{inputenc}
\usepackage{amsmath}
\usepackage{slashed}
\usepackage{dsfont}
\usepackage{amstext}
\usepackage{amssymb}
\usepackage{amsbsy}
\usepackage{amsthm}
\usepackage{epsfig}
\usepackage{graphicx}
\usepackage{bbm}
\usepackage{hyperref}
\usepackage{color}

\begin{document}
 
\title{On the mass enhancement near optimal doping in high magnetic fields in the cuprates}
\author{T. Senthil}
\affiliation{Department of Physics, Massachusetts Institute of Technology,
Cambridge, MA 02139, USA}
 \date{\today}
\begin{abstract}
Recent observation of a mass enhancement in high magnetic fields in nearly optimally doped cuprates poses several puzzles.  For the suggested nodal electron pocket induced by bidirectional charge order in high field, we propose that the mass enhancement is very anisotropic around the small Fermi surface.  The corners of the pocket are proposed to have a big enhancement without any enhancement along the diagonal nodal direction. A natural mechanism for such angle dependent mass enhancement comes from the destruction of the Landau quasiparticle at hot spots on the large Fermi surface at a proximate quantum critical point.  The possibility of a divergent cyclotron effective mass is discussed within a scaling theory for such a quantum critical point.  We consider both a conventional quantum critical point  associated with the onset of charge order in a metal and an unconventional one where an antinodal pseudogap opens simultaneously with the onset of charge order.  The latter may describe the physics of the strange metal regime at zero magnetic field. 
  
\end{abstract}
\newcommand{\be}{\begin{equation}}
\newcommand{\ee}{\end{equation}}
\newcommand{\bea}{\begin{eqnarray}}
\newcommand{\eea}{\end{eqnarray}}
\newcommand{\p}{\partial}
\newcommand{\lp}{\left(}
\newcommand{\rp}{\right)}
\maketitle

It is by now well established\cite{louisT07,qorev}  that quantum oscillations are seen when the superconducting state is suppressed by a magnetic field in the underdoped cuprates. The oscillation frequency points to a small pocket occupying about 2\% of the Brillouin zone for $YBCO_{6.5}$.  This should be contrasted with the Fermi surface of the overdoped cuprates\cite{hussey2003,dmsclli,husseythqo} which occupies more than 60\% of the Brillouin zone. 
Recently Ramshaw et al\cite{rmshw14} reported quantum oscillations in high magnetic field (up to $100 T$) in underdoped YBCO in a doping range approaching optimal doping 
($YBCO_{6+ \delta}$, $\delta = 0.75, 0.80, 0.86$). They found that the oscillation frequency is roughly the same as in the more underdoped samples studied earlier.   However the quasiparticle effective mass $m^*$ extracted from the temperature dependence of the oscillations is strongly enhanced, increasing by a factor $\approx 3$ with increasing doping. Extrapolation of the measured $m^*$ suggests that it diverges at a doping $x \approx 0.18$. This is roughly the same doping level at which many other probes find a change\cite{tllnlrm} of the many body ground state plausibly attributable to a quantum critical point. For instance the charge order reported\cite{hudson08cdw,julien2011cdw,ghiringhelli2012,chang2012cdw,leboeuf2013cdw} in underdoped samples first onsets at this doping\cite{canosaco14}, as does the Kerr effect\cite{kerr}. Further low temperature ARPES\cite{inna12pnas} and STM\cite{seamus14} spectra change significantly around this doping level. 

The results of Ref. \onlinecite{rmshw14} pose several puzzles. Conventional wisdom based on zero field experiments holds that there is no divergent effective mass in the strange metal normal state. For example, ARPES measurements\cite{arpesrev} of Fermi velocity $v_F$ in  the `normal' state above $T_c$ do not see a suppression as optimal doping is approached. Such suppression is also absent  at the $d$-wave Dirac node in the low-$T$ superconducting state both in old synchrotron data\cite{arpesrev} and in the new laser ARPES data\cite{innaprl09,inna12pnas,dessauvf10,uchidandlvf}. Further zero field transport or penetration depth measurements  also do not see any signatures of a divergent effective mass  around optimal doping (for a review see for instance Ref. \onlinecite{lwnrmp06}).

Thus it appears that the mass enhancement of Ref. \onlinecite{rmshw14} is a feature of the metallic state exposed by the high magnetic field. This state has been intensely studied experimentally at lower doping in the last few years\cite{qorev}. The small oscillation frequency is plausibly due to an electron pocket\cite{epckt} located near the nodal region as suggested by Harrison and Sebastian\cite{nhses12,sesnature2014}. This requires bidirectional charge order to reconstruct the large Fermi surface. If the amplitude of the charge order is weak then antinodal pockets will be produced in addition to the electron pocket. For more underdoped samples some other mechanism is needed to get rid of these antinodal pockets and produce the pseudogap which presumably survives at the fields at which quantum oscillations are first seen. 
 
 One of the appealing features of the proposal of Ref. \onlinecite{nhses12} is that the high field metallic Fermi surface essentially inherits the nearly gapless electronic states near the $d$-wave node of the zero field superconductor. If however these near nodal states are seen empirically\cite{inna12pnas,dessauvf10,uchidandlvf} to not have a $v_F$ suppression at $B= 0$, how do they acquire such a suppression when a field is turned on? 
 
Finally why does the observed $m^*$ appear to diverge at the magical doping of $x \approx 0.18$? What, if any, is the relationship between the diverging $m^*$ and the onset of the charge order? 
 
\begin{figure}
\includegraphics[width=2.9 in]{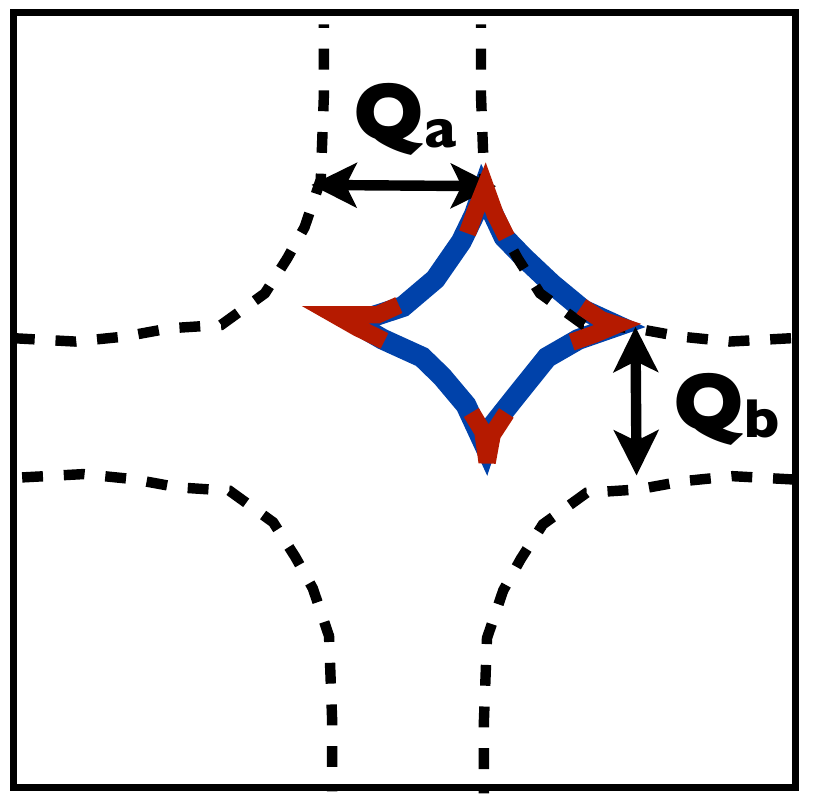}
\caption{Schematic picture of the Harrison-Sebastian nodal electron pocket in the first Brillouin zone. The regions with enhanced effective mass are shown in red. }
\label{smllpckt}
\end{figure}

Here we suggest a resolution of some of these puzzles. Though this is  not crucial, we take for granted the existence of the Harrison-Sebastian electron pocket near the node. We first propose that the mass enhancement is highly anisotropic around this small Fermi surface revealed by the high field. It is enhanced near the corners of this electron pocket (see Figure. \ref{smllpckt}) but not near the node. This leaves the near-nodal states untouched. They can be more or less unmodified from the $B \approx 0$ state. Such a highly angle dependent $m^*(\theta)$  (where $\theta$ is angle around the Fermi surface) can lead to the heavy cyclotron effective mass that is measured in the quantum oscillation experiments. Specifically the quantum oscillation mass $m^*_{QO}$ is obtained as the average
 \begin{equation}
 m^*_{QO} = \frac{1}{2\pi} \oint  \frac{dK}{v_F(\theta)}
 \end{equation}
 where $ v_F(\theta)$ is the Fermi velocity  at angle $\theta$  and $dK$ is the momentum space line element along the Fermi surface.  If there is a local enhanced mass $m^*(\theta)$ that occupies a wide enough portion of the Fermi surface it will dominate the average and lead to an enhanced $m^*_{QO}$. Consider for instance a situation where $v_F(\theta) = v_{Fc}$ in a region of the Fermi surface of  width $\Delta K$ near the corners of the electron pocket. The contribution from the corners to $m^*_{QO}$ will then be of order $\frac{ \Delta K}{v_{Fc}}$.  If this gets enhanced with increasing doping there will be an enhancement of $m^*_{QO}$. 
 
Such an anisotropic $m^*(\theta)$ will not lead to a mass enhancement in transport experiments even in the high field state. Transport will be dominated by the `light' quasiparticles with the highest $v_F$, {\em i.e} the near nodal states.

An independent corroboration of our proposal exists. Recently resistivity measurements have been reported\cite{husseyunpub} in $YBa_2Cu_4O_8$ in the high field regime at low-$T$ (where quantum oscillations are seen) .  The temperature dependence of the resistivity is seen  to be  $A T^2$ consistent with Fermi liquid theory. In many correlated Fermi liquids the $A$ coefficient is typically proportional to $(m^*)^2$ (the `Kadowaki-Woods' plot).  In $YBa_2Cu_4O_8$, $m^*_{QO}$ can also be extracted at high field from the quantum oscillations. However, assuming only a single Fermi surface pocket per Cu-O layer, the measured $A$ is smaller\cite{husseyunpub} by a factor of 20 from that expected from $m^*_{QO}$. This smallness of the resistivity fits nicely with our proposal that the mass enhancement is anisotropic on the Fermi surface so that transport is dominated by the light portions while quantum oscillations is dominated by the heavy portions. 
 
What is a mechanism that can give such an angle dependent effective mass enhancement? We now argue that it is a very natural consequence of proximity to a strange metal with the usual large Fermi surface coupled to critical fluctuations of the density wave order parameter.   At such a quantum critical point there will be ``hot spots" on the large Fermi surface - points that are connected by the charge ordering wave vector at criticality: $\vec Q_a = (Q, 0)$ and $\vec Q_b = (0, Q)$. It is natural that at criticality the Landau quasiparticle is destroyed at these hot spots, and is accompanied by a diverging effective mass (see below). On entering the ordered CDW state the corners of the electron pocket are precisely near these hot spots.  The enhanced corner effective mass is then a signal of the impending loss of the Landau quasiparticle at the hot spots of the large Fermi surface at the proximate quantum critical point.

 To illustrate this mechanism  
we begin first with the conventional model\cite{Hertz} for  the onset of  charge order in a Fermi liquid metal. It assumes that the critical properties can be described adequately by coupling the electronic excitations near the large Fermi surface to fluctuations of the charge density wave order parameter.

To make our considerations precise consider the structure of the electron Green's function in the vicinity of a hot spot (with momentum $\vec K_H$) near criticality. 
This will satisfy a scaling form 
\begin{equation}
G(\vec K, \omega; \Delta) = \frac{c_0}{|\omega|^{\frac{\alpha}{z}} }g\left(\frac{\omega}{c_1 k_\parallel^z}, \frac{k_\perp}{k_\parallel}, \frac{\omega}{\Delta}\right)
\label{ghertz}
\end{equation}
Here $\omega$ is the frequency and $\vec K - \vec K_H = (k_\parallel, k_\perp)$ is the momentum deviation from the hotspot with components perpendicular and parallel to the Fermi surface. $\Delta$ is an energy scale that measures the deviation from the quantum critical point. $z$ is the dynamic critical exponent and is universal as is the exponent $\alpha$. The function $g$ itself will be a universal function while the numbers $c_0, c_1$ are non-universal.  A scaling form of this sort (at $\Delta = 0$) was described in Ref. \onlinecite{maxss10} for the closely analogous problem of the onset of spin density wave order in a metal.  Here we apply this to the CDW onset transition and generalize to $\Delta \neq 0$ as is needed in the present context. 

At $\Delta = 0$, {\em i.e} right at the QCP, when $\frac{k_\perp}{k_\parallel} \rightarrow 0$  we are approaching the Fermi surface away from the hotspot. We then expect to recover Fermi liquid behavior. Matching Eqn. \ref{ghertz} to the Greens function for a Landau quasiparticle 
\begin{equation}
G_{qp} = \frac{Z}{\omega - v_F k_\perp}
\end{equation}
we get the critical properties of $v_F$ (and of $Z$) as the hotspot is approached on the Fermi surface. This leads to $v_F \sim |k_\parallel|^{z-1}$.  Thus if $z > 1$ the Fermi velocity will vanish as the hotspot is approached. 

When $\Delta \neq 0$, the vanishing of the $v_F$ will be cut-off at a momentum scale $|k_\parallel| \sim \Delta^{\frac{1}{z}}$.  This sets the scale $\Delta K $ near the corners of the electron pocket over which the mass enhancement appears. Within this region the Fermi velocity is $v_{Fc} \sim \Delta^{1 - \frac{1}{z}}$. 
Based on our earlier consideration we now see that the singular corner contribution to the quantum oscillation mass is 
\begin{equation}
m^*_{QO} \sim  \Delta^{\frac{2}{z} - 1}
\end{equation}
 So long as $z > 2$ this will diverge and hence so will the measured effective mass in quantum oscillations. 

In the naive Hertz-Millis\cite{Hertz,millisqcp} description of this transition the dynamical exponent $z = 2$ and hence there is no divergence in $m^*_{QO}$ (ignoring log corrections). However modern theoretical work\cite{abchub01,maxss10} shows that the Hertz-Millis theory is incorrect in $d = 2$, and the true strongly coupled theory of this QCP is expected to have a $z$ that is different from $2$. It is currently not known what the critical exponents really are at the transition. We note that a recent study\cite{ssl14} of the analogous spin density wave ordering transition using a novel co-dimensional expansion found an enhancement of $z$ at leading order. 

We can also turn things around and say that within this interpretation, the divergence of $m^*_{QO}$ is experimental proof that $z > 2$ at the CDW onset transition. 

The conventional Hertz model considered so far provides a mechanism illustrating our proposed resolution of the puzzles posed in the beginning. However there is reason to question the direct application of this Hertz model to the cuprates. 

\begin{figure}
\includegraphics[width=2.9 in]{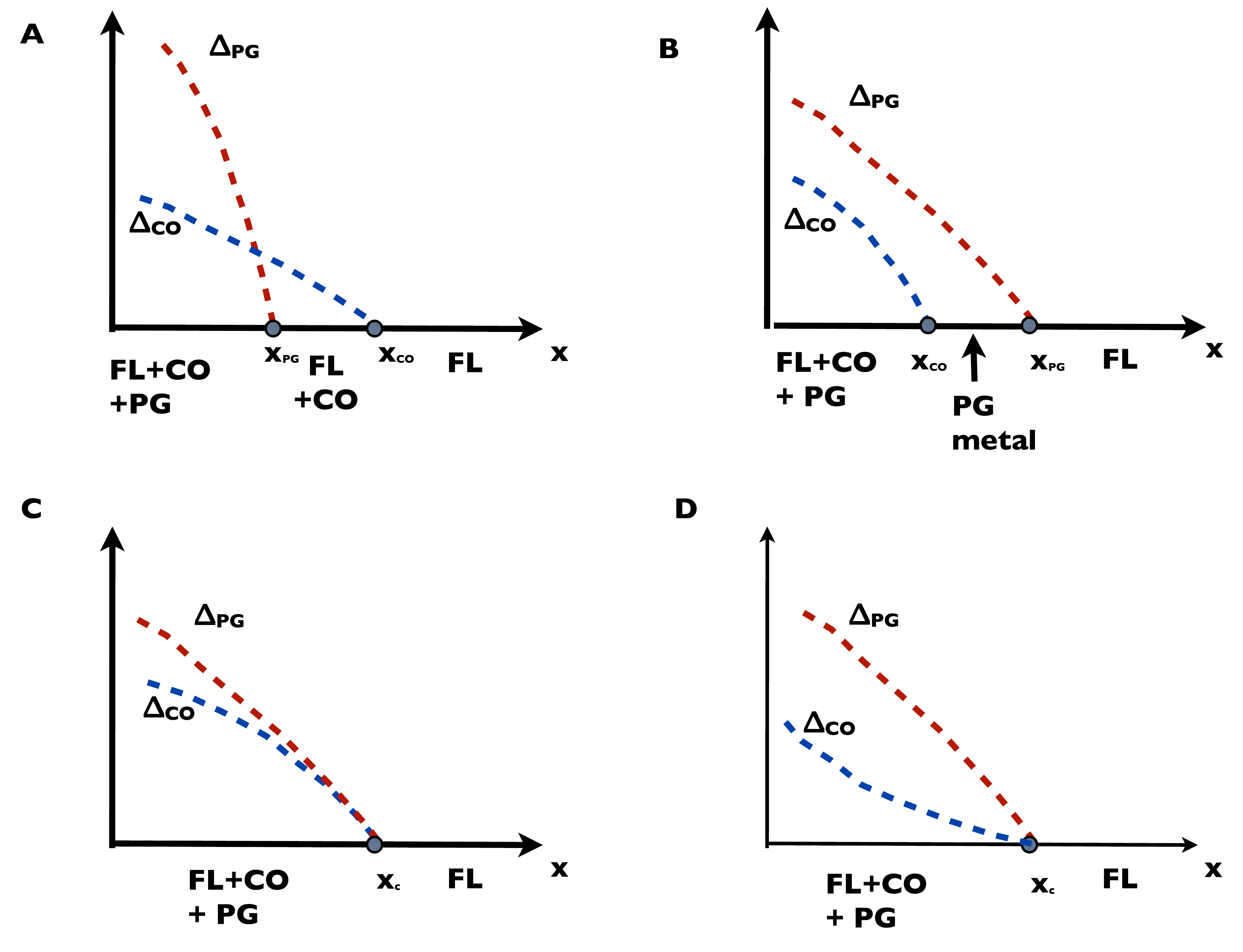}
\caption{Possible evolution of the antinodal pseudogap energy $\Delta_{PG}$ and the energy scale of the CDW ordering $\Delta_{CO}$  (for instance the CDW stiffness) as a function of doping. Only (A) has completely conventional ground states and phase transitions. In (B) the PG metal phase is necessarily unconventional. Both (C) and (D) have conventional phases but the quantum phase transition (if second order)  is unconventional.  }
\label{pdcopg}
\end{figure}

First in this model at the QCP the quasiparticle is destroyed only at the hotspots (a set of measure zero on the Fermi surface). The suggests that this QCP does not directly control the strange metal region of the $B = 0$ phase diagram where the quasiparticle is apparently destroyed over the entire Fermi surface.  Second across the phase transition  apart from the nodal electron pocket there will also be a number of smaller pockets near the antinodal region. This requires that the pseudogap of the underdoped cuprates be destroyed in high field at least near the critical doping $\approx 0.18$. Consequently the high field phase diagram potentially has two different phase transitions as doping is decreased - one associated with the development of charge order, and a subsequent one associated with the opening of the antinodal pseudogap (see Fig. \ref{pdcopg}A). 

Let us consider quite generally how the charge order and the antinodal pseudogap could evolve with doping in an ideal clean system.  To frame the issue sharply let us consider various pertinent energy scales at zero temperature (assuming the superconductivity has been suppressed). The antinodal pseudogap  defines an energy scale $\Delta_{PG}$ which is zero in the overdoped side, and is non-zero well into the  underdoped side.  The charge order may be characterized by an energy scale    $\Delta_{CO}$ which may be taken to be the zero temperature stiffness of the corresponding order parameter.  With increasing $x$ both $\Delta_{PG}$ and $\Delta_{CO}$ need to vanish.  There are 4 general possibilities sketched in Fig. \ref{pdcopg}.  

We emphasize that unlike at non-zero temperature the vanishing of $\Delta_{PG}$ is a genuine $T = 0$ phase transition associated with a change of electronic structure. In the conventional model described above (Case A)  $\Delta_{PG}$ vanishes at a lower critical doping than $\Delta_{CO}$.  Then all the ground states are conventional, the onset of charge order at $x_{CO}$  is described by the Hertz model, and at  $x_{PG}$ there is a Lifshitz transition associated with the disappearance of the antinodal pockets. In Case B, $\Delta_{CO}$ vanishes before $\Delta_{PG}$.   Then there is an  intermediate $T = 0$ metallic phase which has a pseudogap without any broken translation symmetry.  Such a phase is necessarily exotic ( a non-Fermi liquid, in a strict sense). 

Cases (C) and (D) occur if both $\Delta_{PG}$ and $\Delta_{CO}$ vanish at the same doping level.  In (C) they vanish identically ({\em i.e} with the same exponent) on approaching the QCP.  In (D) however they vanish with different exponents \footnote{In Fig. \ref{pdcopg} $\Delta_{CO}$ is assumed to vanish faster than $\Delta_{PG}$. The opposite - where $\Delta_{CO}$ vanishes slower than $\Delta_{PG}$ - can also be envisaged but  seems unlikely to us}.  In both (C) and (D) the ground state phases are not exotic. However the quantum phase transition, if second order,  is very unconventional. The Fermi surface evolves discontinuously across this phase transition. In case (D) near the QCP the charge order may be regarded as a low energy instability of a pseudogap state (itself controlled by an unstable $T = 0$ fixed point; see Ref. \onlinecite{dcss14} for a recent example).

Before proceeding we briefly summarize the experimental situation on the question of whether the pseudogap and charge ordering transitions occur at the same doping level or not. 
There is a considerable literature\cite{tllnlrm} pointing to the termination of the pseudogap around the doping $\approx 0.19$ at zero field, very close to the critical doping\cite{canosaco14} associated with the charge order onset.  Very recent STM experiments\cite{seamus14} show that the locus of $k$-space points where there are coherent Bogoliubov quasiparticles  evolves very rapidly across this doping range. For lower doping this locus defines an arc which roughly resembles the Fermi arc reported by ARPES in the pseudogap normal state. For $x > 0.19$ this locus corresponds to the normal state large Fermi surface. 

Motivated by this we consider now the interesting possibility that the pseudogap and the charge order onset at the same critical doping (Case C or D). The corresponding quantum critical point is {\em not} described by the Hertz model described above but rather requires other fluctuations which lead to the pseudogap, and the concomitant discontinuous evolution of the Fermi surface. The resulting critical fixed point might conceivably control the strange metal regime at $B = 0$.  Then we expect that the Landau quasiparticle is destroyed everywhere on the large Fermi surface at this critical point. 

We will study such a putative critical point through scaling arguments\cite{critfs}  and describe criteria for the corner contribution to the cyclotron effective mass to diverge in the charge ordered pseudo gapped underdoped side.   An effective field theory of this proposed quantum critical point is not currently available and we will not attempt to provide one in this paper. Further we specialize\footnote{We expect Case (D) to behave similarly.} to case (C) which makes the standard assumption that there is only one vanishing energy scale at the QCP (which we commonly denote $\Delta \sim \Delta_{PG} \sim \Delta_{CO}$).  
 
First we expect that at this kind of QCP there is a sharp critical Fermi surface\cite{critfs} without Landau quasiparticles co-existing with critical CDW fluctuations which couple to hot spots on the Fermi surface.  As described in Ref. \onlinecite{critfs}, the electron Green's function near the Fermi surface will satisfy a scaling form. In the present context we expect that the scaling  at the hotspot is different from that near the rest of the Fermi surface. Away from the hotspots we  have\cite{critfs}:
\begin{equation}
G(\vec K, \omega; \Delta) =  \frac{c_0}{|\omega|^{\frac{\alpha}{z}} }g\left(\frac{\omega}{c_1 k^z}, \frac{\omega}{\Delta}\right)
\label{critfsscal}
\end{equation}
Though we continue to use the same notation for the exponents $\alpha, z$ as in Eqn. \ref{ghertz}, they now have a different meaning. Here they describe the scaling as the Fermi surface is approached at a generic point away from the hotspot. $k$ is the deviation of the momentum from the Fermi surface in the direction normal to it. $c_0, c_1$ are non-universal while $\alpha, z$ as well as the scaling function $g$ will be universal. The exponents $\alpha, z$ may, a priori, be angle dependent\cite{critfs} but that will play no role in our discussion.

For $\vec K$ close to a hotspot wave vector $\vec K_H$, we expect a modified scaling form
\begin{equation}
G_{HS}(\vec K, \omega; \Delta) = \frac{\tilde{c_0}}{|\omega|^{\frac{\alpha_H}{z_H}} }g_H \left(\frac{\omega}{\tilde{c_1}k_\parallel^{z_H}}, \frac{k_\perp}{k_\parallel}, \frac{\omega}{\Delta}\right)
\end{equation}
where as before $(k_\parallel, k_\perp)$ are the components of $\vec K - \vec K_H$ in the direction parallel and perpendicular to the Fermi surface respectively. The exponents $z_H, \alpha_H$ are universal but will be distinct from those that describe other portions of the Fermi surface.  
Consider $\vec K$ away from the hotspots. As argued in Ref. \onlinecite{critfs}, since $m^*$ does not diverge at a generic Fermi surface point  on approaching the strange metal, we must have $z = 1$. Thus we can write 
\begin{equation}
G(\vec K, \omega; \Delta) =  \frac{c_0}{|\omega|^{\alpha}}
g\left(\frac{\omega}{v_F k}, \frac{\omega}{\Delta}\right)
\end{equation}
Here we have replaced $c_1$ from Eqn. \ref{critfsscal}  by $v_F$ as it has dimensions of velocity and may be taken to be the Fermi velocity in the limit that the overdoped Fermi liquid approaches the strange metal quantum critical point.

Now consider the scaling form near the hotspot. When $\frac{k_\perp}{k_\parallel} \rightarrow 0$, this must match to the off-hotspot scaling form.  This requires that 
\begin{equation}
g_H(x, y \rightarrow 0, t) \sim x^{\frac{\alpha_H}{z_H} - \alpha}g\left(\frac{x}{y},t\right)
\end{equation}
It follows that at a point of the Fermi surface close to but not at the hotspot,  the electron Green's function behaves as 
\begin{equation}
G(\vec K, \omega; \Delta) =  |k_\parallel|^{\alpha z_H - \alpha_H}\frac{\tilde{c_0}}{|\omega|^{\alpha} }g\left(\frac{\omega}{\tilde{c_1}|k_\parallel|^{z_H - 1}k_\perp}, \frac{\omega}{\Delta}\right)
\end{equation}
This leads to a Fermi velocity $v_F \sim |k_\parallel|^{z_H - 1}$. 

In the charge ordered side this singularity will be cut-off at $|k_\parallel| \sim \Delta^{\frac{1}{z_H}}$. Therefore as before we have for the contribution to $m^*_{QO}$ from the corners of the Harrison-Sebastian pocket
\begin{equation}
m^*_{QO} \sim \Delta^{\frac{2}{z_H} - 1}
\end{equation}
This will diverge so long as $z_H > 2$.  This establishes the promised criterion for this type of quantum criticality to explain the experiments.

The ideas presented here lead to several directions for future theoretical and experimental work. First a better understanding of the conventional density wave transition is clearly called for.  Some of the difficulties in existing theoretical treatments are due to the coupling between  critical density wave and pairing fluctuations. In the high field case discussed in this paper the pairing fluctuations are suppressed and this may enable progress. Second, despite strong suggestive evidence in experiments for the  unconventional critical point discussed above we currently know very little about it theoretically (or even if it can at all exist). Our description incorporates critical charge order fluctuations into an earlier scaling description of such a quantum critical point.  Going beyond the scaling description is a frontier challenge not attempted in this paper. 

An important target for future experiments is to determine which of the four possibilities of Fig. \ref{pdcopg} is realized in the cuprates.  Only case (A) is truly conventional, and requires the presence of a charge ordered phase without an antinodal pseudo gap.   If this is realized in high field,  it will be difficult to relate the charge order onset QCP to the physics of the zero field strange metal. Case (B) is by far the most exotic as it requires the intermediate $T = 0$ pseudogap metal phase without translation symmetry breaking.  Case (D)  is somewhat similar in that such a phase exists but only at intermediate energy scales and is described by a fixed point 
that is eventually unstable to charge ordering.   Case (D) is strongly reminiscent of the phenomenon of deconfined quantum criticality\cite{deccp,deccplong} discussed in spin systems.    To disentangle these possibilities, it will be useful in experiments to study the pseudogap in the $c$-axis transport in high field as a way to get information on the antinodal gap and compare its doping evolution with that of the charge order.

This work was sparked by extensive discussions with Suchitra Sebastian and Mohit Randeria at the Aspen Winter Conference on ``Beyond Quasiparticles: New Paradigms for Quantum Fluids", January 2014. I  am grateful to Nigel Hussey for sharing his unpublished data, and to P. A. Lee and I. Vishik for many useful discussions. 
This work was supported by Department of Energy DESC-8739- ER46872, and partially supported by a Simons Investigator award from the Simons Foundation.

\bibliography{bibl}

\end{document}